\documentclass[prl,twocolumn,amsmath]{revtex4}
\usepackage{graphicx}
\usepackage{dcolumn}
\usepackage{bm}

\begin{document}

\title{Experimental Scheme to Test the Fair Sampling Assumption in EPR-Bell Experiments}

\author{Guillaume Adenier}
\altaffiliation{Pre-doc fellowship of EU-Network ``QP and Applications"}
\email{guillaume.adenier@msi.vxu.se}
\author{Andrei Yu. Khrennikov}%
\email{andrei.khrennikov@msi.vxu.se}
\affiliation{%
Department of Mathematics and System Engineering, V\"axj\"o
University, 351 95 V\"axj\"o, Sweden.
}%

\date{August 26, 2003}

\begin{abstract}
We propose here an experimental test of the fair sampling
assumption in two-channel EPR-Bell experiments for which the
detection loophole holds.
\end{abstract}

\maketitle

In order to assess that a violation of Bell inequality
\cite{bell64} has been experimentally observed, EPR-Bell
experiments \cite{Aspect82,weihs} for which the detection loophole
holds are interpreted assuming \emph{fair sampling}
\cite{Garra,adenkhren}. The purpose of this note is to propose an
experimental scheme to test this fair sampling assumption for
two-channel type experiments.

\begin{figure}
\includegraphics{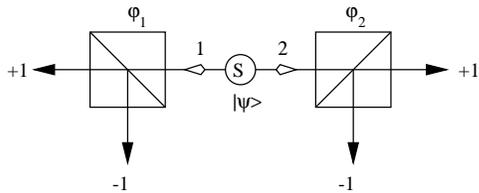}
\caption{\label{fig:epsart8} Two-channel EPR-Bell experiment with
photons. The source is the rotationally invariant singlet state
$|\psi\rangle$ and the measurement is made by two polarizer
beamsplitters with parameters $\varphi_1$ and $\varphi_2$,
monitored by a fourfold coincidence setup (not represented here).}
\end{figure}

Let label as usual the possible measurement results for each
particle as +1 if the particle is detected in the ordinary
channel, as -1 if it is detected in the extraordinary channel, and
as 0 if it remains undetected. Let label the settings of the left
and right two-channel measurement devices respectively as
$\varphi_1$ and $\varphi_2$ (see Fig.~\ref{fig:epsart8}).

With a local realistic model, an apparent violation of a Bell
Inequality can be obtained because of the existence of the 0
channel. The pairs that must remain unregistered for this purpose
must however not be selected completely randomly, but according to
the context of the two-channel measurement devices encountered by
the particles \cite{adenkhren}, that is, the sampling must be
\emph{unfair} and done according to $\varphi_1$ and $\varphi_2$.
Note that it is a completely local process, the rejection of pairs
mimics nonlocality only because of the coincidence circuitry,
since as long as one particle is undetected the whole pair is
naturally rejected.

The trouble with the one-channel experiment that makes checking of
the fair sampling assumption impossible is that the only
measurement result that can be recorded is the +1 result. Hence,
channels -1 and 0 cannot be distinguished as they correspond both
to a non detection. In the two-channel experiment, this
distinction is made, and all four coincidence rates are
available---not only $R_{++}$ as in the one-channel experiment,
but also $R_{+-}$, $R_{-+}$, and $R_{--}$. It is therefore
possible to compute the total rate of detected pairs as
$R_{d}=R_{++}+R_{+-}+R_{-+}+R_{--}$, which as we shall see can
prove to be a crucial tool to test the fair sampling assumption.

Suppose that such an unfair sampling is indeed at stakes in a
two-channel experiment. Let $R_u$ be the rate of pairs rejected
according to this unfair sampling process, and $R_f$ be the rate
of pairs rejected according to a fair one. The total rate $R$ of
pairs entering in the coincidence circuitry is then
$R=R_d+R_f+R_u$. The unfair sampling part $R_u$ is a coherent
process depending strongly on the polarization distribution of the
source and of the orientations $\varphi_1$ and $\varphi_2$ of the
two-channel devices \cite{adenkhren}. On the contrary, the fair
sampling part $R_f$ should be completely independent of these
factors, provided that the rate $R$ of pairs entering in the
coincidence circuitry is a constant of time.

\begin{figure}
\includegraphics[width=8cm]{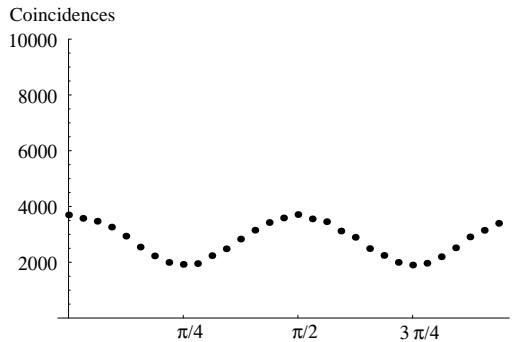}
\caption{\label{fig:epsart6} Numerical simulation of the total
coincidence rate for a source controlled with aligned polarizers
(see Ref. \cite{adenkhren}).}
\end{figure}

Let us assume the source is indeed stable, so that both $R$ and
$R_f$ are invariants for different settings of the instruments. It
means that $R_d+R_u$ must remain invariant too, so that $R_d$ can
be an indirect way to check whether an unfair sampling $R_u$ is at
stakes. Since it is however possible to build a local realistic
model that exhibit independent errors \cite{Larsson2}, in which
$R_d$ remains a constant for a source of entangled photons,
whatever $\varphi_1$ and $\varphi_2$, and since a model with
dependent errors exhibit in this case very small oscillations of
$R_d$ \cite{adenkhren} anyway, observing that $R_d$ is invariant
for a source of entangled photon cannot be taken as a proof that
there is no unfair sampling at stakes. Hence, our proposal to test
fair sampling for the two-channel experiment is to modify and
control the source. Instead of the rotationally invariant singlet
state, we propose to send a set of pairs already sampled, i.e.,
with a preferred direction of polarization $\theta$. Provided that
there is an unfair sampling at stakes with the measurement setup,
then when this $\theta$-sampled source meets the unfair sampling
measurement setup depending on $\varphi$, $R_u$ and therefore
$R_d$ should exhibit some variations when $\theta$ is varied. The
result of simulations \cite{adenkhren} shows a strong variation
(see Fig.~\ref{fig:epsart6}).

\begin{figure}
\includegraphics[width=8.5cm]{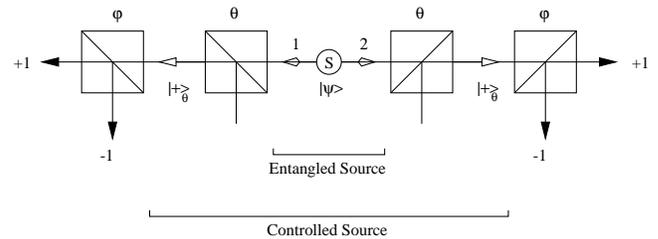}
\caption{\label{fig:epsart7} Scheme of controlled EPR-Bell
experiment to test fair sampling. The only difference with the
usual EPR-Bell experiment (see Fig.~\ref{fig:epsart8}) is the
source, which is controlled by the parameter $\theta$. In case of
unfair sampling, this setup should exhibit oscillations for the
total coincidence rate as in Fig.~\ref{fig:epsart6}.}
\end{figure}

In order to control the source experimentally, our proposal is
thus to insert aligned polarizers oriented along an angle \(\theta
\) in the coincidence circuitry before the two-channel measurement
devices, both oriented along the same angle \(\varphi \) (see
Fig.~\ref{fig:epsart7}). If the sampling is fair then $R_d$ should
remain invariant when the angle \(\theta \) is varied with respect
to \(\varphi \), and no such oscillations as in
Fig.~\ref{fig:epsart6} should be observed.

The fair sampling test we propose here should be simple to
implement, and should allow to either discard all known local
realist models based on detection loophole, or on the contrary, to
show that the analyzers which are assumed to be ideal two-channel
measuring devices are in fact unfair and therefore inappropriate
for an EPR-Bell test.

\bibliography{proposal}

\begin{thebibliography}{6}
\expandafter\ifx\csname natexlab\endcsname\relax\def\natexlab#1{#1}\fi
\expandafter\ifx\csname bibnamefont\endcsname\relax
  \def\bibnamefont#1{#1}\fi
\expandafter\ifx\csname bibfnamefont\endcsname\relax
  \def\bibfnamefont#1{#1}\fi
\expandafter\ifx\csname citenamefont\endcsname\relax
  \def\citenamefont#1{#1}\fi
\expandafter\ifx\csname url\endcsname\relax
  \def\url#1{\texttt{#1}}\fi
\expandafter\ifx\csname urlprefix\endcsname\relax\def\urlprefix{URL }\fi
\providecommand{\bibinfo}[2]{#2}
\providecommand{\eprint}[2][]{\url{#2}}

\bibitem[{\citenamefont{Bell}(1964)}]{bell64}
\bibinfo{author}{\bibfnamefont{J.~S.} \bibnamefont{Bell}},
  \bibinfo{journal}{Physics} \textbf{\bibinfo{volume}{1}}, \bibinfo{pages}{195}
  (\bibinfo{year}{1964}).

\bibitem[{\citenamefont{Aspect et~al.}(1982)\citenamefont{Aspect, Dalibard, and
  Roger}}]{Aspect82}
\bibinfo{author}{\bibfnamefont{A.}~\bibnamefont{Aspect}},
  \bibinfo{author}{\bibfnamefont{J.}~\bibnamefont{Dalibard}}, \bibnamefont{and}
  \bibinfo{author}{\bibfnamefont{G.}~\bibnamefont{Roger}},
  \bibinfo{journal}{Phys. Rev. Lett.} \textbf{\bibinfo{volume}{49}},
  \bibinfo{pages}{91} (\bibinfo{year}{1982}).

\bibitem[{\citenamefont{Weihs et~al.}(1998)\citenamefont{Weihs, Jennewein,
  Simon, Weinfurter, and Zeilinger}}]{weihs}
\bibinfo{author}{\bibfnamefont{G.}~\bibnamefont{Weihs}},
  \bibinfo{author}{\bibfnamefont{T.}~\bibnamefont{Jennewein}},
  \bibinfo{author}{\bibfnamefont{C.}~\bibnamefont{Simon}},
  \bibinfo{author}{\bibfnamefont{H.}~\bibnamefont{Weinfurter}},
  \bibnamefont{and}
  \bibinfo{author}{\bibfnamefont{A.}~\bibnamefont{Zeilinger}},
  \bibinfo{journal}{Phys. Rev. Lett.} \textbf{\bibinfo{volume}{81}},
  \bibinfo{pages}{5039} (\bibinfo{year}{1998}).

\bibitem[{\citenamefont{Adenier and Khrennikov}()}]{adenkhren}
\bibinfo{author}{\bibfnamefont{G.}~\bibnamefont{Adenier}} \bibnamefont{and}
  \bibinfo{author}{\bibfnamefont{A.}~\bibnamefont{Khrennikov}},
  \eprint{quant-ph/0306045}.

\bibitem[{\citenamefont{Garrucio and Rapisarda}(1981)}]{Garra}
\bibinfo{author}{\bibfnamefont{A.}~\bibnamefont{Garrucio}} \bibnamefont{and}
  \bibinfo{author}{\bibfnamefont{V.}~\bibnamefont{Rapisarda}},
  \bibinfo{journal}{Nuovo Cimento} \textbf{\bibinfo{volume}{A65}},
  \bibinfo{pages}{269} (\bibinfo{year}{1981}).

\bibitem[{\citenamefont{Larsson}(2000)}]{Larsson2}
\bibinfo{author}{\bibfnamefont{J.~A.} \bibnamefont{Larsson}}, Ph.D. thesis,
  \bibinfo{address}{Link\"oping Univ. Press, Sweden} (\bibinfo{year}{2000}).

\end{thebibliography}

\end{document}